\documentclass[%
reprint,
 amsmath,amssymb,
 aps,
]{revtex4-2}

\usepackage{graphicx}
\usepackage{dcolumn}
\usepackage{bm}
\usepackage[section]{placeins}
\usepackage{color}
\usepackage{mathrsfs}
\usepackage{ulem}
\usepackage{hyperref}
\usepackage{braket}
\usepackage{siunitx}

\begin{document}

\title{Wide-field NV magnetometry under simultaneous high-pressure and high-temperature conditions}

\author{Masahiro Ohkuma$^{1}$}
\email{okuma.m.408d@m.isct.ac.jp}
\author{Eikichi Kimura$^{1}$}
\author{Shumpei Ohyama$^{1}$}
\author{Miu Tezuka$^{2}$}
\author{Ryo Matsumoto$^{3}$}
\author{Shinobu Onoda$^{4}$}
\author{Yoshihiko Takano$^{3}$}
\author{Shintaro Azuma$^{2}$}
\author{Kenji Ohta$^{2}$}
\email{k-ohta@eps.sci.isct.ac.jp}
\author{Keigo Arai$^{1}$}
\email{arai.k.835f@m.isct.ac.jp}
\affiliation{$^{1}$School of Engineering, Institute of Science Tokyo, Yokohama 226-8501, Kanagawa, Japan}
\affiliation{$^{2}$Department of Earth and Planetary Sciences, Institute of Science Tokyo, Meguro 152-8551, Tokyo, Japan}
\affiliation{$^{3}$Research Center for Materials Nanoarchitectonics (MANA), National Institute for Materials Science, Tsukuba 305-0047, Ibaraki, Japan}
\affiliation{$^{4}$National Institutes for Quantum Science and Technology (QST), Takasaki 370-1292, Gunma, Japan}

\date{\today}

\begin{abstract}
We demonstrate wide-field optically detected magnetic resonance (ODMR) under simultaneous high-pressure and high-temperature conditions using nitrogen-vacancy (NV) centers.
Although NV-center magnetometry has been widely used for spatially resolved magnetic-field imaging, its application to extreme environments combining pressure and temperature remains challenging.
In this work, we show that ODMR can be observed at 5 GPa and 500 K, demonstrating the feasibility of NV spin readout under such combined extreme conditions.
We further perform wide-field ODMR of iron at 7 GPa and 500 K, where the stray magnetic field from the sample is spatially visualized through the pressure cell.
These results establish NV-center magnetometry as a promising platform for imaging magnetic phenomena in materials under high-pressure and high-temperature environments.
\end{abstract}

\maketitle

\section {Introduction}
High pressure and high temperature are fundamental thermodynamic variables for controlling the structural, electronic, and magnetic states of materials and have also been widely used to synthesize functional materials that are inaccessible under ambient conditions \cite{shenHighpressureStudiesXrays2016, maoSolidsLiquidsGases2018, yamanakaSiliconClathratesCarbon2010, azumaFunctionalTransitionMetal2021}.
In high-pressure experiments using diamond anvil cell (DAC), simultaneous high-pressure and high-temperature conditions are commonly achieved by combining the DAC with heating techniques such as external resistive heating, internal resistive heating, or laser heating \cite{hazenHightemperatureDiamondanvilPressure1981, liuMeltingIron2001975, mingLaserHeatingDiamond1974}.
These developments have enabled a wide range of in situ structural, spectroscopic, and electrical transport measurements under high-pressure and high-temperature conditions \cite{salamatSituSynchrotronXray2014, linSituHighPressuretemperature2004, kantorLaserHeatingFacility2018, yousufHighpressureHightemperatureElectrical1986, ohtaMeasuringElectricalResistivity2023}.
Such measurements provide essential information for understanding phase transitions, magnetic ordering, and functional properties in condensed-matter physics, materials science, and geoscience.

Magnetic measurements under simultaneous high-pressure and high-temperature conditions are still challenging.
M\"ossbauer spectroscopy and X-ray magnetic circular dichroism (XMCD) have been used to investigate the magnetic properties under high pressure and high temperature \cite{pipkornMossbauerEffectIron1964, kantorPressureInducedMagnetizationFeO2004, liMonoclinicDistortionMagnetic2024, mathonDynamicsMagneticStructural2004, dewaeleMagneticPhaseDiagram2022}.
These techniques provide information on hyperfine interactions and element-specific magnetic moments. 
However, M\"ossbauer spectroscopy is restricted to suitable M\"ossbauer-active nuclei, whereas XMCD can be constrained in high-pressure experiments by the accessible X-ray absorption edges and X-ray transmission through the diamond anvils \cite{cranshawMossbauerSpectroscopy1974, mathonXMCDPressureFe2004, haskelInstrumentXrayMagnetic2007}.
In contrast, magnetometry using commercial magnetometor based superconducting quantum interference device (SQUID) systems is a standard and versatile approach for magnetic characterization, however, widely used systems in their standard configuration typically operate only up to approximately 400 K \cite{mitoMagnetizationMeasurementsUsing2025}.
Moreover, magnetic measurements under high pressure are further constrained by the small sample volume in the pressure chamber and the background signals from the pressure cell \cite{mitoMagnetizationMeasurementsUsing2025}.
Therefore, a broadly applicable technique for probing local magnetic fields under simultaneous high-pressure and high-temperature conditions is desirable.

\begin{figure}[htb!]
\centering\includegraphics[clip,width=1\columnwidth]{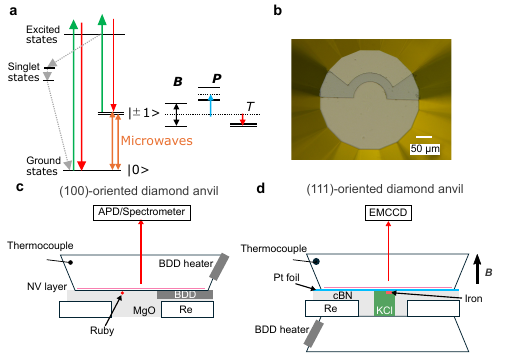}
\caption{\label{f1} Experimental configurations for ODMR measurements under high pressure.
(a) Energy-level diagram of NV center.
(b) Optical image of the diamond anvil with boron-doped diamond antenna.
(c) Schematic of the DAC setup for ODMR measurements using (100)-oriented diamond anvil.
(d) Schematic of the DAC setup for wide-field ODMR measurements using (111)-oriented diamond anvil.}
\end{figure}

The negatively charged nitrogen-vacancy (NV) center in diamond offers a promising route to local magnetic sensing under extreme conditions.
It is a point defect that has enabled versatile applications in quantum information processing and sensing \cite{jelezkoObservationCoherentOscillations2004, gaebelRoomtemperatureCoherentCoupling2006, jelezkoObservationCoherentOscillation2004, hansonPolarizationReadoutCoupled2006, duttQuantumRegisterBased2007, taylorHighsensitivityDiamondMagnetometer2008a, degenScanningMagneticField2008, balasubramanianNanoscaleImagingMagnetometry2008a, mazeNanoscaleMagneticSensing2008, acostaTemperatureDependenceNitrogenVacancy2010, doldeElectricfieldSensingUsing2011, kucskoNanometrescaleThermometryLiving2013, lesageOpticalMagneticImaging2013}.
The NV center has a spin-triplet ground state that can be initialized and read out optically through spin-dependent fluorescence and exhibits magnetic resonance transitions at a zero-field splitting ($D$) of approximately 2.87 GHz under ambient conditions, as shown in Fig.~\ref{f1}a \cite{dohertyNitrogenvacancyColourCentre2013}.
Detection of these magnetic resonance through changes in the fluorescence intensity is known as optically detected magnetic resonance (ODMR).
The energy levels of the state are highly sensitive to external perturbations, such as temperature, magnetic fields, and pressure, thereby enabling nanoscale sensing of these physical quantities \cite{taylorHighsensitivityDiamondMagnetometer2008a, degenScanningMagneticField2008, balasubramanianNanoscaleImagingMagnetometry2008a, mazeNanoscaleMagneticSensing2008, acostaTemperatureDependenceNitrogenVacancy2010, doldeElectricfieldSensingUsing2011, kucskoNanometrescaleThermometryLiving2013, lesageOpticalMagneticImaging2013, dohertyElectronicPropertiesMetrology2014, broadwayMicroscopicImagingStress2019}.
The zero-field splitting $D$ increases under hydrostatic pressure with a pressure coefficient of approximately 14 MHz/GPa, whereas it decreases with increasing temperature at a rate of approximately $-74$ kHz/K near room temperature
\cite{acostaTemperatureDependenceNitrogenVacancy2010, dohertyElectronicPropertiesMetrology2014, hoSpectroscopicStudyN$V$2023}.
In contrast, a magnetic field $B$ applied along the NV axis lifts the degeneracy of the $m_s=\pm1$ states through the Zeeman effect, shifting the resonance frequencies by approximately $\pm \gamma_{\mathrm{e}}B_{\parallel}$, where gyromagnetic ratio $\gamma_{\mathrm{e}}\approx28$ MHz/mT.
Therefore, the frequency separation between the two ODMR resonances is approximately $2\gamma_{\mathrm{e}}B_{\parallel}$.
The NV center retains its spin coherence even under temperatures up to 1400 K \cite{toyliMeasurementControlSingle2012, liuCoherentQuantumControl2019, fanQuantumCoherenceControl2024} or pressure above 100 GPa \cite{hsiehImagingStressMagnetism2019, lesikMagneticMeasurementsMicrometersized2019, yipMeasuringMagneticField2019, bhattacharyyaImagingMeissnerEffect2024, wangImagingMagneticTransition2024}.
Although high-pressure or high-temperature NV-magnetometry have been demonstrated separately, NV-based magnetic imaging under simultaneous high-pressure and high-temperature conditions remains largely unexplored.

In this study, we demonstrate wide-field ODMR simultaneous high-pressure and high-temperature conditions.
Using local resistive heating with a patterned boron-doped diamond heater, we observe ODMR signals from NV centers at 5 GPa and 500 K, confirming that NV spin readout can be maintained in this environment.
We further perform magnetic imaging of iron under 7 GPa and 500 K, and visualize the stray magnetic field from the sample.
These results extend NV-magnetometry to spatially resolved magnetic imaging across pressure--temperature phase space.

\section {Methods}
Pressure was applied using a CuBe diamond anvil cell. 
We used type-Ib single-crystal diamond anvils with (100)- and (111)-oriented culet surfaces (Syntek Co., Ltd.). 
The (100)-oriented diamond anvil was used for ODMR measurements, whereas the (111)-oriented diamond anvil was used for wide-field ODMR of iron.
The culet diameter of both diamond anvils was 300 $\mu$m.
For the creation of NV centers, vacancies were introduced by ${}^{12}\mathrm{C}^{+}$ ion implantation at an energy of 30 keV with a fluence of $5\times10^{12}$ cm$^{-2}$. 
The diamond anvils were then annealed under vacuum at \SI{1000}{\degreeCelsius} for 2 h. 
Stopping and Range of Ions in Matter simulations indicate that the implanted vacancies are distributed within approximately 50 nm from the diamond surface \cite{zieglerSRIMStoppingRange2010}.

To heat the sample space, we employed a boron-doped diamond (BDD) heater 
\cite{takanoSuperconductivityDiamondThin2004,matsumotoNoteNovelDiamond2016,matsumotoDiamondAnvilCell2021,ohkumaCoherentControlSolidstate2024}. 
For the (100)-oriented diamond-anvil experiment, the BDD heater was fabricated on the NV-containing diamond anvil. 
For the (111)-oriented diamond-anvil experiment, the BDD heater was fabricated on the opposing type-IIa diamond anvil. 
Microwave excitation was delivered using a BDD antenna patterned on the culet surface for the (100)-oriented diamond-anvil experiment, whereas a Pt foil was used for microwave delivery in the (111)-oriented diamond-anvil experiment.
BDD circuits were fabricated after the formation of NV centers.
An optical image of the (100)-oriented diamond anvil is shown in Fig.~\ref{f1}b.

A rhenium gasket was pre-indented and drilled to form a hole with a diameter of 100 or 250 $\mu$m. 
For the (100)-oriented diamond-anvil experiment, MgO was used to electrically insulate the BDD antenna from the metallic gasket. 
For the (111)-oriented diamond-anvil experiment, a cBN--epoxy insulating layer was compressed, and a 100-$\mu$m-diameter hole was subsequently drilled. 
KCl was used as the pressure-transmitting medium.
In the wide-dield ODMR experiment using the (111)-oriented diamond anvil, an iron foil was loaded into the sample chamber.
The temperature was monitored using a K-type thermocouple attached to the slope of the NV-containing diamond anvil with cement. 
The diamond anvil was mounted on a zirconia seat and fixed with cement.
For the (100)-oriented diamond-anvil experiment, the pressure was estimated from ruby fluorescence 
\cite{weiFluorescencePressureSensors2024}.
For the (111)-oriented diamond-anvil experiment, the pressure was estimated from resonance frequency of [111]-oriented NV center \cite{maiMegabarPressureSensing2025}. 
An overview of the experimental setups is illustrated in Figs.~\ref{f1}c and \ref{f1}d, where Fig.~\ref{f1}c shows the setup for ODMR measurements and Fig.~\ref{f1}d shows the setup for wide-field ODMR.

Continuous-wave (CW) ODMR measurements were performed using a custom-built microscope. 
A 532-nm green laser (MLL-S-532B, CNI laser or Verdi 2G, Coherent) was used to excite the NV centers through an objective lens (M-PLAN APO SL 20X, Mitutoyo). 
The red fluorescence from the NV centers was collected through the same objective lens and passed through a long-pass filter. 
For point-detection ODMR measurements, the fluorescence was detected by an avalanche photodiode (APD130A2, Thorlabs) coupled to a data acquisition device (USB-6363, National Instruments). 
For wide-field ODMR imaging, the fluorescence was detected by an EMCCD camera (iXon Ultra 897, Andor).
Microwaves were generated using a signal generator (N5171B, Keysight, or SynthHD, Windfreak) and amplified using a power amplifier (ZHL-50W-63$+$ or ZHL-16W-43$+$, Mini-Circuits).
A point-detection ODMR spectrum was acquired within 2 min, whereas a wide-field ODMR spectrum was acquired within 15 min.

\section{Results}
\begin{figure}[htb!]
\centering\includegraphics[]{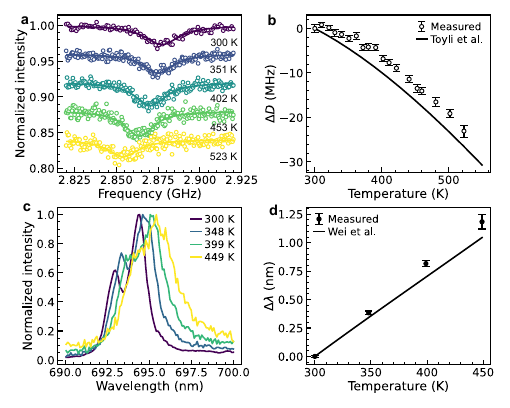}
\caption{\label{f2} (a) CW-ODMR spectra of NV centers measured at different temperatures.
The ODMR spectra are vertically offset by 0.04 for clarity.
(b) Temperature dependence of the zero-field splitting shift, $\Delta D = D(T)-D(300~\mathrm{K})$.
The solid line represents the reported temperature dependence of zero-field splitting shift from Ref.~\cite{toyliMeasurementControlSingle2012}.
The measured shift is in reasonable agreement with the reported temperature dependence.
(c) Ruby fluorescence spectra measured at different temperatures. 
(d) Temperature dependence of the R1-line wavelength shift, $\Delta\lambda=\lambda(T)-\lambda(300~\mathrm{K})$.
The solid line represents the reported temperature dependence of R1-line wavelength shift from Ref.~\cite{weiFluorescencePressureSensors2024}.
The agreement of both the NV and ruby temperature dependences confirms that the thermocouple attached to the diamond anvil provides a reliable measure of the temperature in the present experimental configuration.}
\end{figure}

To verify the reliability of the temperature measured using the thermocouple attached to the slope of the diamond anvil, we measured the temperature dependence of the ODMR spectra and the ruby fluorescence spectra at ambient pressure.
Figures~\ref{f2}a and \ref{f2}b show the ODMR spectra measured at different temperature and the corresponding temperature dependence of the zero-field splitting $D$, respectively.
Here, $D$ is plotted as a function of temperature $T$ relative to its value at 300 K, $\Delta D = D(T)-D(300~\mathrm{K})$.
The relatively low apparent ODMR contrast in this measurement may be due to the non-optimized optical collection conditions, which allowed part of the ruby fluorescence to be collected together with the NV fluorescence.
The measured temperature dependence of $\Delta D$ is in reasonable agreement with the previous report \cite{toyliMeasurementControlSingle2012}.
Figures~\ref{f2}c and \ref{f2}d show the ruby fluorescence spectra measured at different temperatures and the corresponding temperature dependence of the R1 fluorescence line wavelength, respectively. 
The R1 line shift is plotted relative to its value at 300 K, $\Delta\lambda = \lambda(T)-\lambda(300~\mathrm{K})$. 
The observed temperature dependence is also consistent with previous study. 
These results confirm that the thermocouple attached to the diamond anvil provides a reliable measure of the temperature in the sample chamber.

\begin{figure}[htb!]
\centering\includegraphics[]{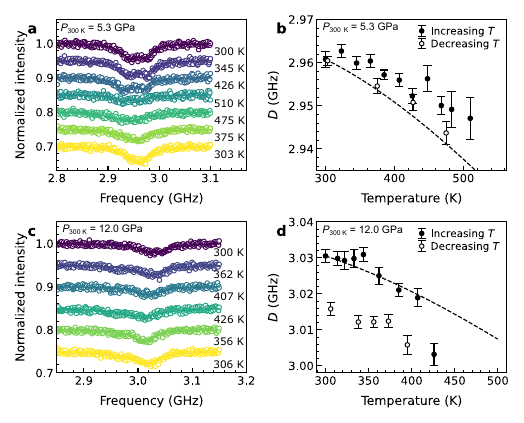}
\caption{\label{f3} ODMR measurements under simultaneous high-pressure and high-temperature conditions.
(a) CW-ODMR spectra of NV centers measured at different temperatures at an initial pressure of 5.3 GPa before heating.
Magnetic resonance signals were observed even above 500 K.
(b) Temperature dependence of the zero-field splitting parameter $D$ for the data shown in (a).
The dashed line represents a temperature-only reference curve calculated from the polynomial temperature dependence of $D$ reported by Toyli et al. \cite{toyliMeasurementControlSingle2012}, after shifting the curve to match the experimentally measured value of $D$ at 300 K.
(c) CW-ODMR spectra of NV centers measured at different temperatures at an initial pressure of 12.0 GPa before heating.
Magnetic resonance signals were observed up to 426 K.
(d) Temperature dependence of $D$ for the data shown in (c).
The dashed line represents the temperature-only reference curve obtained in the same manner as in (b).
The ODMR spectra are vertically offset by 0.05 for clarity.}
\end{figure}

The results obtained under applied pressure are summarized in Fig.~\ref{f3}.
Figure~\ref{f3}a shows the ODMR spectra measured at an initial ruby pressure of 5.3 GPa before heating.
Clear magnetic resonance signals were observed even above 500 K, demonstrating that ODMR readout is possible under simultaneous high-pressure and high-temperature conditions.
Figure~\ref{f3}b shows the temperature dependence of $D$.
The dashed line represents a temperature-only reference curve calculated from the polynomial temperature dependence reported by Toyli $et~al.$ \cite{toyliMeasurementControlSingle2012}, with the curve offset to match the measured value of $D$ at 300 K.
The temperature dependence of $D$ under pressure qualitatively follows the trend expected from the thermal shift.
However, the measured $D$ values remain higher than those expected from the temperature-induced shift alone.
This deviation suggests that change in pressure also contributes to the observed resonance shift during heating.
Ruby fluorescence measurements performed after the temperature-dependent ODMR measurements indicated that the pressure had increased from 5.3 to 6.1 GPa.
After cooling back to room temperature, however, $D$ returned to nearly the same value as that before heating, whereas ruby fluorescence indicated an increase in pressure after the heating cycle.
This discrepancy is likely due to the difference in the local pressure and stress environments probed by the NV centers and the ruby particle. 
The NV centers are located within approximately 50 nm of the diamond anvil surface, whereas the ruby particle is placed in the sample chamber.

Figure~\ref{f3}c shows the ODMR spectra measured at an initial ruby pressure of 12.0 GPa before heating.
Magnetic resonance signals were observed up to 426 K, confirming that ODMR readout remains possible even at higher pressure.
Figure~\ref{f3}d shows the temperature dependence of $D$.
Up to approximately 400 K, the temperature dependence of $D$ follows the trend expected from the thermal shift.
At 426 K, however, $D$ decreases abruptly.
This abrupt decrease is consistent with a pressure decrease during heating, as supported by ruby fluorescence measurements after the temperature-dependent ODMR measurements, which showed that the pressure had decreased from 12.0 to 10.4 GPa.
These results demonstrate that ODMR can be observed under simultaneous high-pressure and high-temperature conditions.
At the same time, they indicate that accurate in situ pressure calibration at elevated temperature is required to quantitatively determine the pressure--temperature dependence of the $D$.

Next, we performed wide-field ODMR measurements under simultaneous high-pressure and high-temperature conditions.
Before discussing the wide-field ODMR results, we describe how the pressure was estimated in the wide-field ODMR experiment.
The pressure was estimated from the temperature-corrected shift of $D$ in the (111)-oriented diamond anvil \cite{maiMegabarPressureSensing2025}.
For simplicity, we assumed that the temperature- and pressure-induced shifts of $D$ can be treated independently.
The temperature-induced shift from 300 K, denoted as $\Delta D_T(T)=D_{\mathrm{ref}}(T)-D_{\mathrm{ref}}(300~\mathrm{K})$, was calculated from the reported temperature dependence of $D$ at ambient pressure \cite{toyliMeasurementControlSingle2012}.
The pressure was then estimated as
\begin{equation}
P_{\mathrm{ODMR}}(T)=
\frac{D_{\mathrm{meas}}-\Delta D_T(T)-D_0}{\kappa},
\label{eq:P_ODMR}
\end{equation}
where $D_{\mathrm{meas}}$ is the measured zero-field splitting, $D_0=2.87~\mathrm{GHz}$ is the zero-field splitting at 300 K and ambient pressure, and $\kappa=7.94~\mathrm{MHz/GPa}$ is the pressure coefficient for the (111)-oriented diamond anvil \cite{maiMegabarPressureSensing2025}.
Because the pressure was estimated using the temperature correction reported at ambient pressure, the obtained value should be regarded as an approximate pressure near the NV layer.

\begin{figure}[htb!]
\centering\includegraphics[]{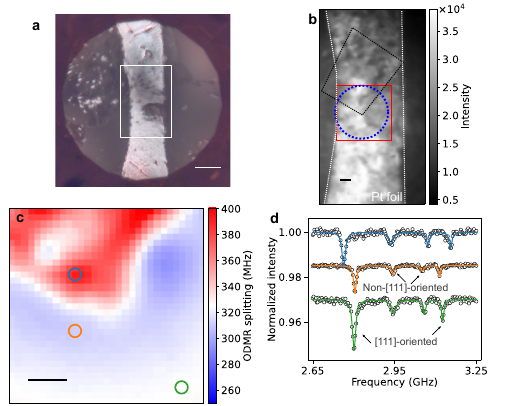}
\caption{\label{f4} Wide-field ODMR at $P_{\mathrm{ODMR}}=11.8$ GPa and 300 K.
(a) Optical scope image inside diamond anvil cell. The scale bar indicates 50 $\mu$m.
(b) Wide-field fluorescence image in the region of interest (ROI) shown in (a).
The sample chamber position was identified from an optical image taken from the opposite side of the NV-containing diamond anvil. 
The black dotted line shows the estimated outline of the iron foil, the red solid square indicates the ROI used for ODMR analysis, and the blue dotted circle indicates the KCl-filled sample chamber region.
(c) ODMR splitting map obtained from pixel-by-pixel fitting of the wide-field ODMR spectra in the ROI shown in (b). 
The ODMR splitting is defined as the frequency separation between the two resonance dips assigned to the [111]-oriented NV centers under an applied bias magnetic field of approximately 5.7 mT.
Enhanced and reduced splitting regions correspond to stray-field components from the iron foil that are parallel and antiparallel to the applied bias field along the [111] NV axis, respectively.
The colored circles indicate the positions at which the representative ODMR spectra in (c) were extracted.
(d) Representative ODMR spectra obtained from regions with enhanced splitting, reduced splitting, and far from the iron foil.
The ODMR spectra are vertically offset by 0.015 for clarity.
The scale bars in (b) and (c) represent 10 $\mu$m.}
\end{figure}

Figure~\ref{f4}a and b show a optical scope image and a wide-field fluorescence image acquired at $P_{\mathrm{ODMR}}=11.8$ GPa and 300 K.
At the pressure--temperature conditions studied here, iron is expected to remain in the ferromagnetic region according to the reported magnetic phase diagram, although $\alpha$-Fe transforms to nonferromagnetic $\epsilon$-Fe at around 13--15 GPa under hydrostatic compression at room temperature \cite{dewaeleMagneticPhaseDiagram2022}.
The position of the sample chamber was identified from an optical image taken from the opposite side of the NV-containing diamond anvil.
Only part of the iron foil was directly visible; therefore, the overall foil position was inferred from the visible portion, the known foil geometry, and the ODMR-derived magnetic-field map.
The region of interest used for ODMR analysis was defined to encompass the KCl-filled sample chamber region.
A bias magnetic field of approximately 5.7 mT was applied using a permanent magnet, with the field direction approximately aligned along the [111] direction of the diamond anvil. 
We define the ODMR splitting as the frequency separation between the two resonance dips assigned to the (111)-oriented NV centers.

Figure~\ref{f4}b shows the spatial map of ODMR splitting obtained from the wide-field ODMR spectra.
The observed splitting map contains both enhanced and reduced regions relative to the background splitting far from the sample. 
This behavior can be understood as the spatial variation of the stray magnetic field from the iron foil projected along the [111] NV axis. 
Where the stray-field component is parallel to the applied bias field, the total field along the NV axis increases and the ODMR splitting becomes larger. 
In contrast, where the stray-field component is antiparallel to the bias field, the total field is partially cancelled and the ODMR splitting becomes smaller. 
To illustrate the spectral origin of the contrast in the splitting map, representative ODMR spectra obtained from regions with enhanced splitting, reduced splitting, and far from the iron foil are shown in Fig.~\ref{f4}c.

\begin{figure}[htb!]
\centering\includegraphics[]{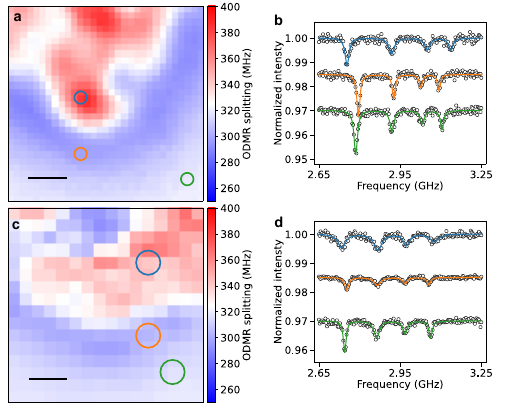}
\caption{\label{f5} Wide-field ODMR under simultaneous high-pressure and high-temperature conditions.
(a) ODMR splitting map measured at 430 K and $P_{\mathrm{ODMR}}=11.2$ GPa.
(b) Representative ODMR spectra obtained from regions with enhanced splitting, reduced splitting, and far from the iron foil for the data shown in (a).
(c) ODMR splitting map measured at 500 K and $P_{\mathrm{ODMR}}=7.2$ GPa.
Camera binning was employed in this measurement to improve the signal-to-noise ratio within the limited acquisition time, resulting in coarser spatial sampling.
(d) Representative ODMR spectra obtained from regions with enhanced splitting, reduced splitting, and far from the iron foil for the data shown in (c).
The ODMR spectra are vertically offset by 0.015 for clarity.
The scale bars in (a) and (c) represent 10 $\mu$m.}
\end{figure}

Next, we increased the temperature from the condition shown in Fig.~\ref{f4} and performed wide-field ODMR measurements under simultaneous high-pressure and high-temperature conditions.
Figure~\ref{f5}a shows the ODMR splitting map measured at 430 K and $P_{\mathrm{ODMR}}=11.2$ GPa.
A spatial variation of the ODMR splitting is still observed near the iron foil, indicating that the stray magnetic field from the sample can be imaged under high-pressure and high-temperature conditions.
Representative ODMR spectra obtained from regions with enhanced splitting, reduced splitting, and far from the iron foil are shown in Fig.~\ref{f5}b.

Figures~\ref{f5}c and \ref{f5}d show the corresponding results measured at 500 K and $P_{\mathrm{ODMR}}=7.2$ GPa.
In this measurement, camera binning was employed to improve the signal-to-noise ratio within the limited acquisition time, resulting in a coarser spatial sampling than in the lower-temperature measurements.
The pressure estimated from the ODMR shift decreased during heating, possibly caused by deformation of the sample chamber.
Indeed, the sample chamber had been deformed after the heating experiment.
A clear spatial variation of the ODMR splitting remains visible at higher temperature.
These results demonstrate that NV-based wide-field magnetic imaging can be performed at temperatures above 500 K under applied pressure.

\begin{figure}[htb!]
\centering\includegraphics[]{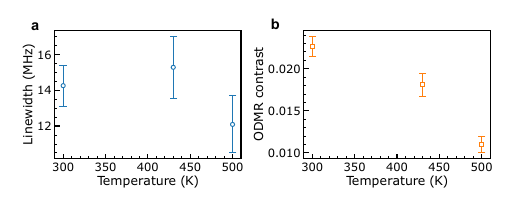}
\caption{\label{f6} Temperature dependence of ODMR spectral parameters under high-pressure and high-temperature conditions.
(a) Linewidth of the lower-frequency resonance assigned to the [111]-oriented NV centers, extracted from ODMR spectra acquired far from the iron foil.
(b) ODMR contrast extracted from the same resonance.
The extraction regions are indicated by the green regions in the corresponding splitting maps.
The linewidth shows only a weak temperature dependence, whereas the ODMR contrast decreases with increasing temperature.}
\end{figure}

Figures~\ref{f6}a and \ref{f6}b summarize the linewidth and ODMR contrast, respectively, as a function of temperature.
These values were extracted from the lower-frequency ODMR resonance assigned to the [111]-oriented NV centers in spectra acquired far from the iron foil, as indicated by the green regions in the corresponding splitting maps.
The linewidth shows only a weak temperature dependence over the measured temperature range.
In contrast, the ODMR contrast decreases with increasing temperature, as previously reported \cite{toyliMeasurementControlSingle2012}.

\section {Discussion}
The present results demonstrate that NV-magnetometry can serve as local magnetic probes under simultaneous high-pressure and high-temperature conditions. 
Compared with conventional bulk magnetometry, the present approach provides spatially resolved information from a sample inside a DAC. 
This capability is complementary to element- or isotope-selective probes such as XMCD and M\"ossbauer spectroscopy, because NV magnetometry detects the local magnetic field generated by the sample. 
Thus, NV-based magnetic imaging may provide a new route for investigating magnetic properties under pressure and temperature.

The temperature dependence of $D$ is known to originate from phonon-mediated effects, and recent analytical models describe $D(T)$ in terms of the thermal occupation of representative phonon modes \cite{dohertyTemperatureShiftsResonances2014, ivadyPressureTemperatureDependence2014, tangFirstPrinciplesCalculationTemperatureDependent2023,cambriaPhysicallyMotivatedAnalytical2023}. 
Under applied pressure, phonon energies generally increase, and therefore the temperature dependence of $D$ may also be modified by pressure. 
However, this effect is expected to be small in diamond at the pressure range studied here because of its lattice stiffness. 
This expectation is supported by high-pressure Raman studies showing only a moderate pressure dependence of the diamond phonon modes
\cite{akahamaHighpressureRamanSpectroscopy2004}.
Assuming that the relevant phonon modes contributing to $D(T)$ have a comparable pressure dependence, the change in the calculated thermal shift of $D$ between 300 and 500 K is expected to be only marginal at around 10 GPa.

There is still considerable scope to extend the experimental range in both temperature and pressure.
For temperature, CW ODMR is applicable up to approximately 700 K, while pulsed-ODMR with laser heating allows measurements up to 1400 K \cite{toyliMeasurementControlSingle2012, liuCoherentQuantumControl2019, fanQuantumCoherenceControl2024}.
In practice, however, repeated short-timescale pressure variations during heating cycles in pulsed-ODMR are expected to limit the experimental feasibility.
Accordingly, the practical temperature limit is likely around 700 K.
Reaching higher pressures will also require experiments under hydrostatic conditions using fluid pressure-transmitting media, because non-hydrostatic pressure reduces the ODMR contrast \cite{hoSpectroscopicStudyN$V$2023,huangElucidatingIntersystemCrossing2026}.

\section {Conclusion}
In conclusion, we demonstrated NV-based magnetic imaging under simultaneous high-pressure and high-temperature conditions. 
ODMR signals from NV centers were observed up to 500 K under applied pressure, and wide-field ODMR measurements visualized the stray magnetic field from an iron foil under the same combined extreme conditions.
These results show that NV-center magnetometry can be extended to spatially resolved magnetic imaging under concurrent high-pressure and high-temperature conditions. 
Further studies on the pressure and temperature dependence of the zero-field splitting under in situ pressure and temperature calibration will be essential for quantitative NV magnetometry in high-pressure and high-temperature environments.

\begin{acknowledgments}
This work was supported by JSPS KAKENHI Grant numbers JP24KJ1035, JP23K26528, JP23KK0267.
This work was also supported by JST ASPIRE Grant number JPMJAP24C1.
M.O. receives funding from JSPS Grant-in-Aid for JSPS Fellows Grant number JP24KJ1035.
E.K. receives funding from JST SPRING, Japan Grant Number JPMJSP2180.
\end{acknowledgments}

\section*{DATA AVAILABILITY}
The data that support the findings of this article are openly available \cite{ohkumaReplicationHPHT2026}.

\bibliography{NV}

\end{document}